# A new facility for High energy PIXE at the ARRONAX Facility

C. Koumeir[1], F. Haddad[1,2], V. Metivier[1], N. Servagent[1] and N. Michel[1,2]

(1) SUBATECH, Université de Nantes, Ecole des Mines de Nantes, CNRS/IN2P3, La Chantrerie, 4, rue A. Kastler, BP 20722, 44307 Nantes, France
(2) GIP ARRONAX, 1 rue Aronnax, Saint Herblain France

## Abstract

Proton Induced X-ray Emission (PIXE) using high energy protons is a non destructive multi elemental technique that can analyze medium and heavy trace elements on thick samples. A new experimental setup is being built at the ARRONAX facility (Nantes, France) for such purpose. Tests have been made in order to quantify the background induced by a 68 MeV proton beam on a 250μm copper target. Our measurement has been compared with expected theoretical value including the various components of the bremsstrahlung and the Compton induced by gamma rays. This study has allowed us to determine a detection limit of the order of tens of ppm and has shown also the different ways to improve it. Amongst them are a better shielding of the detector and the use of lower beam intensity. Another ways consist to use the prompt gamma rays emitted by the interaction of the beam with the target nucleus as well as the registration of the decay of the activation product after the end of the bombardment. This activation and especially the production of long lived radionuclide can be controlled by tuning the proton energy and intensity resulting in a low activation which is not prejudicial to most of the samples and with almost no effect on the PIXE performances.

## High energy PIXE

Proton induced X-ray emission (PIXE) is a non destructive multi elemental technique allowing to determine concentration of elements with $Z > 11$. This method has been applied with success in many different field using low-energy protons (1-5 MeV) with a detection limit in the order of ppm [1]. The thickness of analyzed samples is limited by the range of low-energy protons and is of the order of 35μm for 3 MeV protons in Cu. Analysis is done using K X-rays for light element and L-Xrays for heavy element. It is then difficult sometimes to deconvolve two peaks from neighbour elements [2].

High energy PIXE presents several advantages compared to low energy PIXE. Heavy element can be studied using their K X-ray emission since their production cross-section increase with the proton incident energy (see Fig. 1). This additional information can help analysing the data. Thick target can be studied thanks to the much larger range of the energetic protons and to the smaller absorption of the hard K X-rays. Heavy elements can be detected as far as few millimetres deep inside the sample. For example, the range of 30 MeV protons in copper matrix is ~ 1.6 mm and K X-Rays for Sn (Z = 50) in copper can be seen at a depth of ~ 1mm. Additionally, for high energy protons, the energy loss, the energy straggling and the angular diffusion are low. It is then possible to make experiments directly in air with almost no disturbance of the beam. Data analysis is simpler due to the slow evolution of the X-ray production cross section value and moreover, the low straggling in energy induce more accurate results [3].

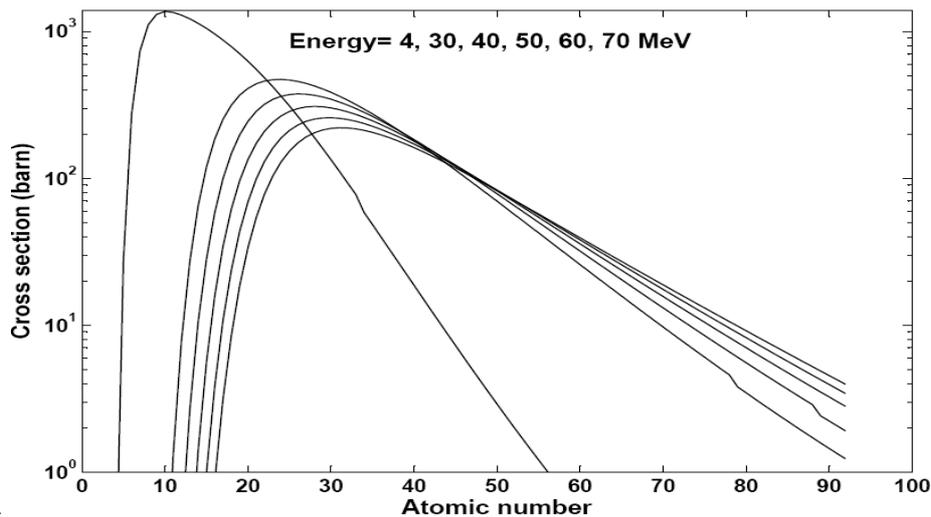

**Figure 1** K X-Ray Cross section for proton incident energy= 4, 30, 40, 50, 60, 70 MeV as a function of the atomic number. These curves were calculated by an analytical cross-section formula [4]

High energy PIXE has already been used in several applications in the past [5, 6, 7, 8]: In Geology, rare earth elements which can be measured easily thanks to the K X-ray emission cross section which is not negligible, are important for the understanding of geological phenomena [5]. In Archaeology, historical metal objects (for example coins) are often covered with corrosion layer (sometimes ~100 µm thick) [6]. Measurements of the composition of the bulk without removing this layer are then possible with HEPIXE. It should also be possible to measure the concentration profile in thick sample to control the diffusion of certain elements in nuclear materials.

Besides the advantages of HEPIXE, there are two identified drawbacks: At high energy, the background level is expected to be important affecting the detection limit and nuclear reaction can occurred leading to the activation of the samples.

In this paper we will present our first results obtained at the ARRONAX facility. A special care has been devoted to the modeling of the background in order to be able to estimate the detection limit of our set-up. During the experiments, gamma rays have been identified and their possible use for on-line (prompt gamma emission) and off-line (activation gamma rays) analysis have been studied. The sample activation can be controlled by adjusting both intensity and proton energy.

**Experimental setup**

ARRONAX, acronym for "Accelerator for Research in Radiochemistry and Oncology at Nantes Atlantique", is a high energy (70 MeV) and high intensity cyclotron (up to 750µA) [9]. It is mainly devoted to the production of radionuclide for medicine. It is a multi-particle accelerator: proton can be accelerated from 30 MeV up to 70 MeV, deuteron from 15 up to 35 MeV and alpha-particle at a fixed 68 MeV. It contains 6 experimental vaults one, called AX, being devoted to research in physics and radiochemistry. Several experimental set-up are being build among which a high energy PIXE. Fig. 2 is a schematic view of the experimental setup. A target made of a 250µm thick copper foil (S=625 mm$^2$) and covered on one side by a 30 µm thick Bismuth layer was used. The Cu foil (purity >99%) was bought from Goodfellow Inc whereas the Bi layer was obtained by thermal deposition under vacuum. Its homogeneity is around 6%. This Bi layer has been added for monitoring purpose. X-rays were collected by a high purity germanium from Canberra. The Ge crystal has a thickness of 5mm and an active area of 50mm². In order to reduce the detection threshold, the detector is mounted with a 50μm Be window. To limit the effect of the background, the detector has been shielded using 5cm thick lead brick. The angle of detection is chosen to be 135° with respect to the beam and the distance between the detector and the target is 105cm to limit the counting rate at a reasonable value (dead time < 10%). The experiment was made in normal air and we used a 68 MeV proton beam and a beam intensity lower than 20 nA (the lowest proton beam intensity on the accelerator at the moment). A typical run duration time is 5mn.

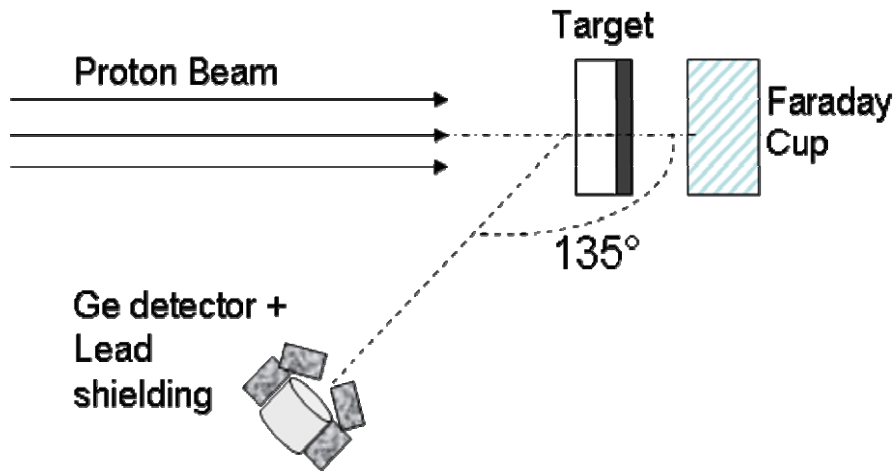

Figure 2: Schematic view of the experiment setup

The gray curve on Fig. 3 shows the background spectra (blank target) whereas the black curve corresponds to an experiment with the Copper/Bismuth target, the copper layer being set downstream the beam line. In both spectra, we can see the X-rays lines coming from the ionization of the lead K shell by the various gamma rays present in the background. With a target in position, other peaks can be observed. They are associated respectively to Cu and Bi. Small peaks can be found around channel 160 and 260. They have been identified as prompt gamma emission associated to $^{nat}Cu(p,x)^{62}Cu$ (gamma emitted at 41 keV) and to $^{nat}Cu(p,x)^{61}Ni$ (gamma emitted at 67,4 keV).

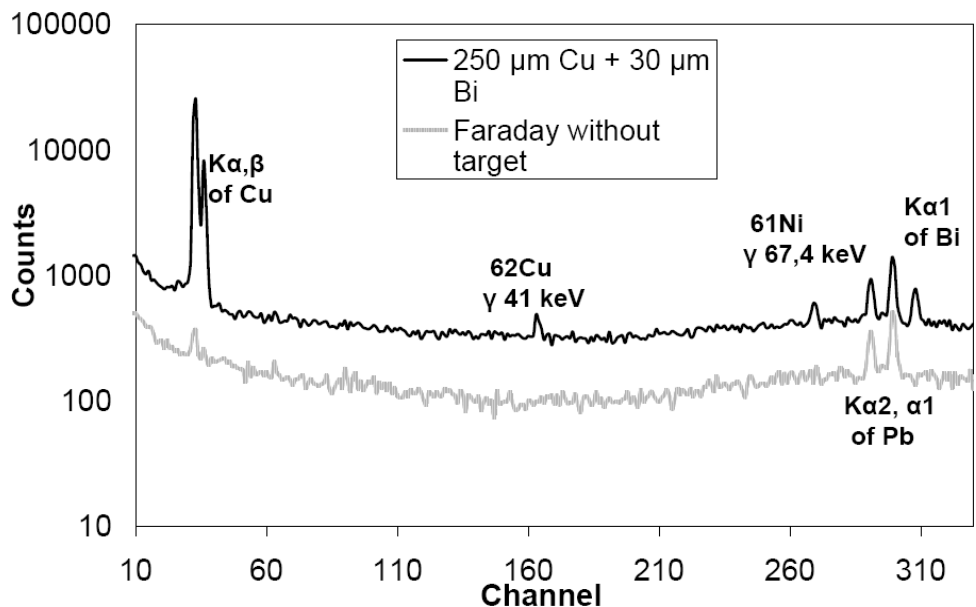

Figure 3 Experimental spectrum measured at 135° for 68 MeV proton

# Target background

To be able to define the detection limit precisely, it is important to understand the background structure of our spectrum. Fig. 4 presents a spectrum recorded during a longer time (500 s). All the other irradiation parameters were the same as Fig. 3.

A blank target measurement has also been done. A simple subtraction after normalisation using the lead K X-ray peaks and a smoothing procedure has then be applied to obtain the background component produced within the target itself (see the curve below the spectrum in Fig. 4). The other sources, interaction of the beam with the Faraday cup and the gamma and neutron from the environment, being eliminate by the previous subtraction. This curve will be used in the following part of the paper to compare with a theoretical estimate of background coming from the target.

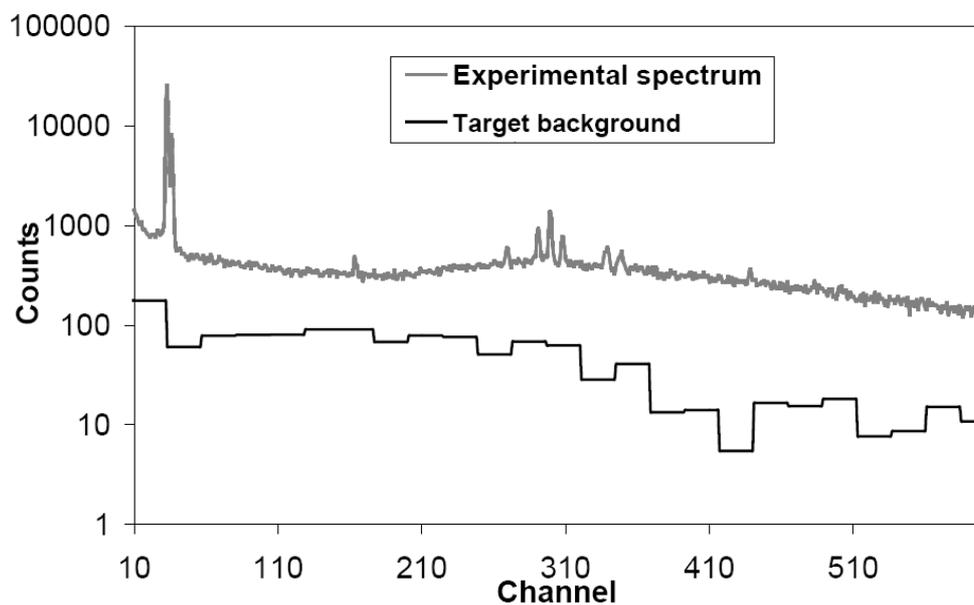

**Figure 4 Spectrum of the target in the presence of the Faraday cup and the curve of the estimated background**

The interaction of protons with the target creates many photons by both electron bremsstrahlung and Compton diffusion of gamma rays. At high energy, the bremsstrahlung is mainly formed by two components: secondary electron bremsstrahlung (SEB), which corresponds to the diffusion of secondary electrons by atoms of the medium, and the quasi electron bremsstrahlung (QFEB) which corresponds to the electron scattering in the projectile frame. The bremsstrahlung is less intense in the backward direction [10] and that is why we chose a detection angle of 135 degrees. Calculation of SEB is based on the binary-encounter approximation while QFEB is based on the plane-wave Born approximation [11].

The Compton is due to gamma rays with energies <3 MeV. This limit is imposed by the thickness of our germanium crystal which is well suited for low energy photons and has a very small efficiency above this value. These gamma rays are emitted by the excited target nuclei formed by the proton interaction on the target nuclei. In order to get a realistic gamma ray spectrum, we use the nuclear physics TALYS code [12]. The Compton component is then calculated using the equations of the Klein-Nishina [13].

Fig. 5 presents calculated spectra: QFEB (dashed line), SEB (dash-dotted line), Compton (long-dashed line) and the sum spectrum (solid line). These spectra were calculated using the same experimental conditions (energy and beam intensity, detection solid angle) for a target formed with 250 μm Cu and 30 μm Bi. Absorption in the target, in the air and within the Be window, as well as the detector efficiency were taken into consideration. We cut our calculated spectra below 10 keV since the bremsstrahlung is very sensitive to absorption at such low energy and calculations become difficult. Generally, the QFEB is limited by the energy of the electron in the coordinate of the proton (~ 34 keV), and SEB by the maximum energy transferred during the front collision between the proton and a free electron (~ 68 keV). But in our case, because both the cross section of the bremsstrahlung emission at 135 ° and the solid angle are low, these limits are not reached. The jumps in the Compton spectrum are due to the beginning of Compton front for each gamma. Experimentally, these jumps are not seen due to the resolution of the detector.

This graph shows that at high incident energy, the Compton component dominates the high energy part of the spectra. This is related to two effects: the first is the increase of the cross section of nuclear reactions, thus increasing the intensity of gamma rays, and the second is the reduction of secondary electrons number generated in the medium (the energy loss is low) thus the decrease in the intensity of the bremsstrahlung (SEB).

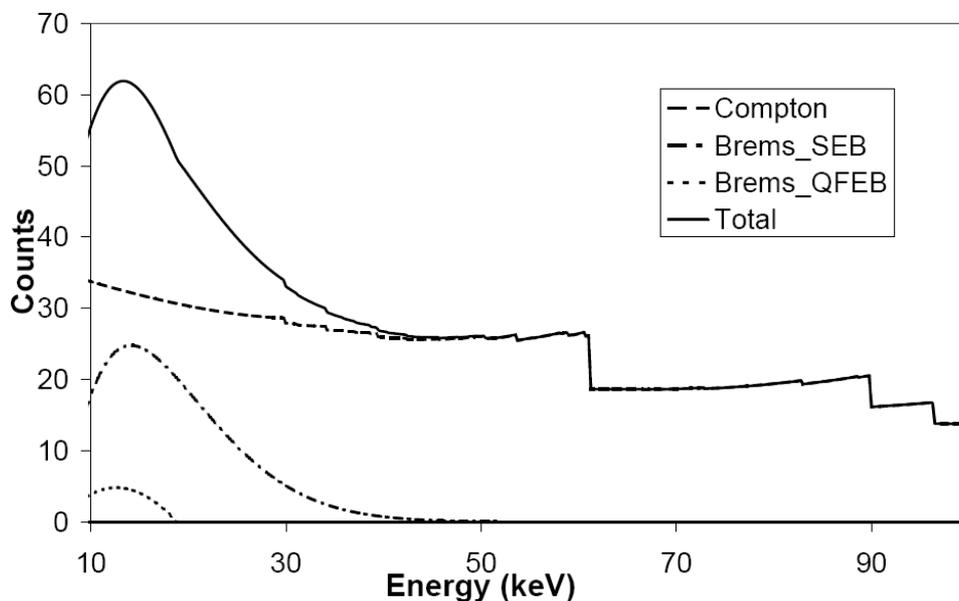

**Figure 5 Compton and bremsstrahlung calculated for proton bombardment of a 250μm Cu +30 μm Bi target at 68 MeV, corrected with absorption in target and air**

Being able to calculate the different background contributions, it is then possible to compare calculated and experimental background. Fig. 6 shows such a comparison. Above 100 keV, where the Compton, component is the most important one, we found a good agreement. Below 100 keV, the theoretical spectrum does not reproduce the experimental data. This disagreement can be due to our subtraction procedure which is intended to eliminate the background coming from the environment. Indeed, it has been found that the environment is changing with the irradiation time. This evolution of background is related to the fact that the beam intensity is high and our shielding was probably not optimum.

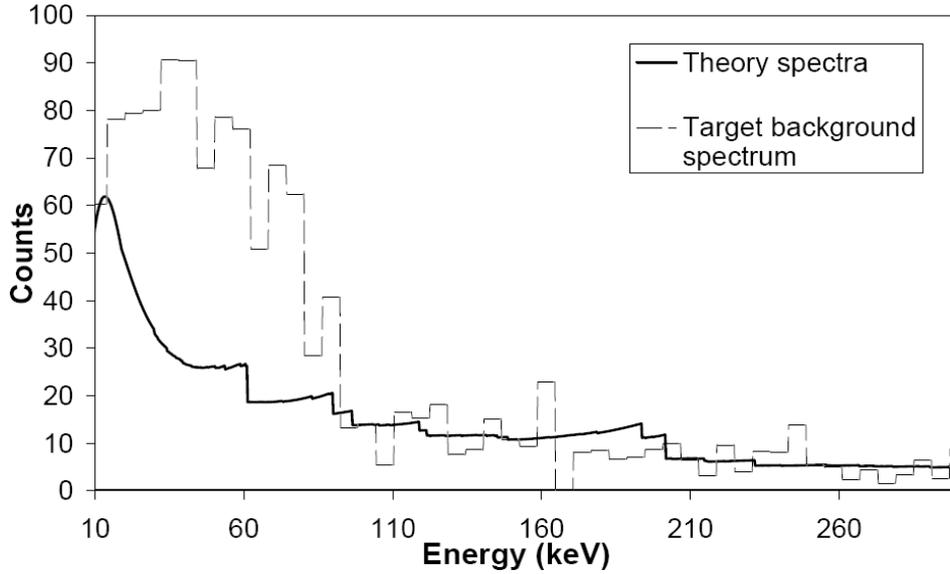
**Figure 6 Comparison between experimental and theoretical background of the target**

## Detection limits

The detection limit is the key parameter for defining the interest of such a method. For a given trace element A with atomic number Z in a given matrix B, the detection limit is determined by statistical fluctuation of background and therefore it is defined by:

$$N_A \geq 3 \times \sqrt{N_B} \quad \text{[a]}$$

where $N_A$ is the total number of counts of a characteristic X-ray peak for a given line i of A, and $N_B$ is the number of background counts included in the full width of half maximum (FWHM) of the characteristic X-ray peak.

The detection limit of A in the matrix B is given in units of parts per million by the following formula [1]:

$$\frac{n_A}{n_B} = \frac{3 \times \sqrt{N_B} \times 10^6}{n_B \times N_P \times \left(\frac{\Omega}{4 \times \pi}\right) \times \sigma_Z^i \times ab(Z) \times \varepsilon_f(Z)} \quad \text{[b]}$$

where $n_B$ is the atomic concentration of the matrix element, $n_A$ is that of the trace element A, $N_p$ is the number of projectiles, $\Omega$ is the solid angle subtended by a detector, and $\sigma_Z^i$, $ab(Z)$, $\varepsilon_f(Z)$ are, respectively, the production cross section of K X-rays for the trace element, absorption of X-rays by windows and others (air, target) and the detection efficiency.

Fig. 7 shows the detection limit calculated with formula [b] for a copper matrix and the already mentioned experimental conditions. In our case, the K X-ray production cross section is quite constant since the energy loss of 68 MeV proton in 250 μm copper is only ~ 1.5 MeV. The experimental background has been used to calculate $N_B$ (dashed line in Fig. 6).

The detection limit varies between 10 and 100 ppm for most elements. It reaches a minimum for Z between 30-40 (the sensitivity is maximum) because the K X-ray production cross section is highest in this region for the incident energy of 68 MeV (Fig. 1).

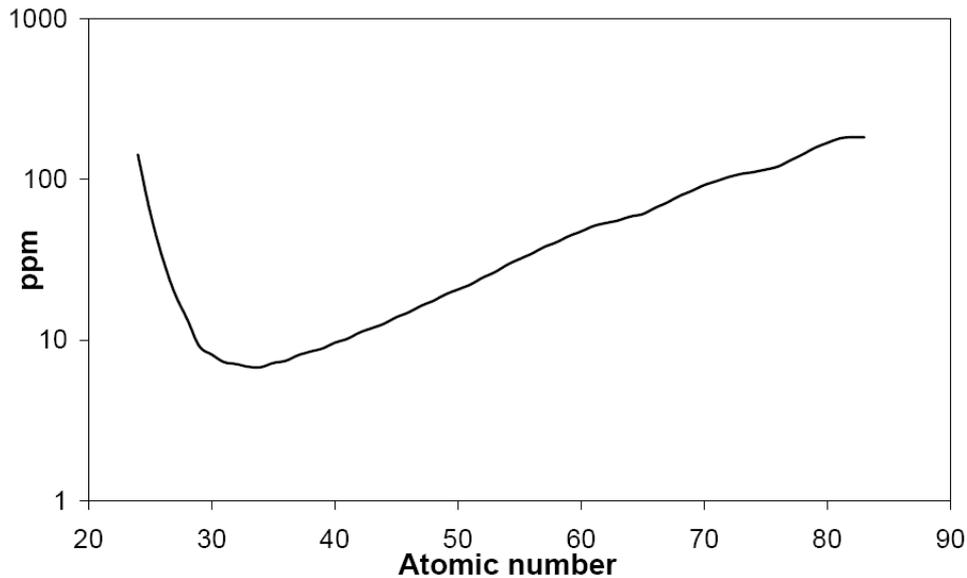
**Figure 7** The detection limit for PIXE analysis in copper matrix (250 µm) by proton (68 Mev) bombardment based on K X-ray detection

These results are encouraging. We expect to reduce further this limit by using a better detector shielding (3 layers (Pb, Cu, Al) compact shielding) and by reducing the beam intensity. We also intend to get additional information from prompt gamma and activation gamma.

## Gamma rays analysis

At high incident energy, nuclear reactions occur and isotopes can be produced within the target. It can be nuclei in excited states (stable or radioactive) that will decay by prompt gamma emission or radionuclide that will decay according to their half lives. For each sample, it is possible to get a good idea of all the produced nuclei by using the TALYS code. In our experiment, the gamma ray peak at 41 keV is associated to prompt gamma emission from formed $^{61}$Ni and the gamma ray peak at 67.4 keV comes from the $^{62}$Cu formed in an exited state. This copper 62 (half life: $T1/2 = 9.673$ mn) will undergo further decay mainly via beta+ emission.

These peaks which are produced by the interaction of the beam with copper can be used to follow the copper deeper inside matter. Indeed, K X-ray of copper are at low energy and the absorption in matter is important limiting the zone of interest close to the surface. Knowing the nuclear data associated to the reaction mechanisms that produce copper, it is possible to determine the detection limit of copper in a thick target (> 200 µm). As an example, at a depth of 200µm, this value is greater than that achieved with X-rays by a factor ~ 2. In the one hand, the production cross section of these gamma rays is lower than that of X-rays by a factor of ~ 5000. On the other hand, at 200µm the absorption coefficient of 67.4 keV gamma rays is lower than Cu K X-rays by a factor ~ 10000. The use of prompt gamma ray may enhance the range of accessible nuclei deep inside thick samples.

Since, most of the produced nuclei from a copper target are unstable (such as zinc, copper, nickel, cobalt, iron and manganese), they will decay by electron capture or β decay, leading to excited daughter nuclide. Among these parent isotopes, there are some who have a period $T_{1/2}$ of the order of minutes and hours. Thus, they can be measured when the beam is turned off using a high precision shielded gamma ray detector. This off line measurement can give additional information about the composition of the target on a macroscopic level helping define the measurement strategy.

## Activation control

When instable nuclei with half lives of days or weeks (long-lived isotopes) are produced, activation can become a problem. In our experiments, we measure the dose on our samples at different times. Two weeks after our experiments, our samples have returned to the background level of radioactivity. This is due to the short half life of the majority of produced isotopes. Indeed, according to a TALYS calculation, only small amount of Cobalt isotopes are produced which are the only long live isotopes in this case with no effect on the sample dose. Nevertheless, it is possible to control to a certain level the activation sample by choosing the incident proton energy as well as limiting the beam intensity. The latter point is obvious since the activation is directly related to the number of incident proton. As an example, in our case, if we reduce the beam intensity down to 50 pA, with the same counting time (500 sec), and target-detector distance of 5 cm, we can register the same spectra that were measured (Fig. 2). In this case, we produce 400 times less radionuclides and we get also a better signal to background ratio.

The other point is related to the shape of the reaction cross section and the fact that most of the nuclear reaction are endothermic and thus can occurred only if a certain incident beam energy is delivered (energy threshold). In addition, K X-ray production cross sections varied only slightly with respect to proton energy (see Fig 1). A small changed in the incident proton energy will then have only a small effect on the HEPIXE performances but a large effect on the sample activation. Fig. 8 displays the production cross section of different Cobalt isotopes using natural Copper target ($^{57}$Co and $^{60}$Co), and the total production cross section (all isotopes) as function of the proton energy. It can be seen that below 35 MeV none of these two isotopes are produced. The total production cross section stays almost constant above 15 MeV. The main sample activation comes from the matrix which is often known. For trace element, activation will be small and its effect can be neglected.

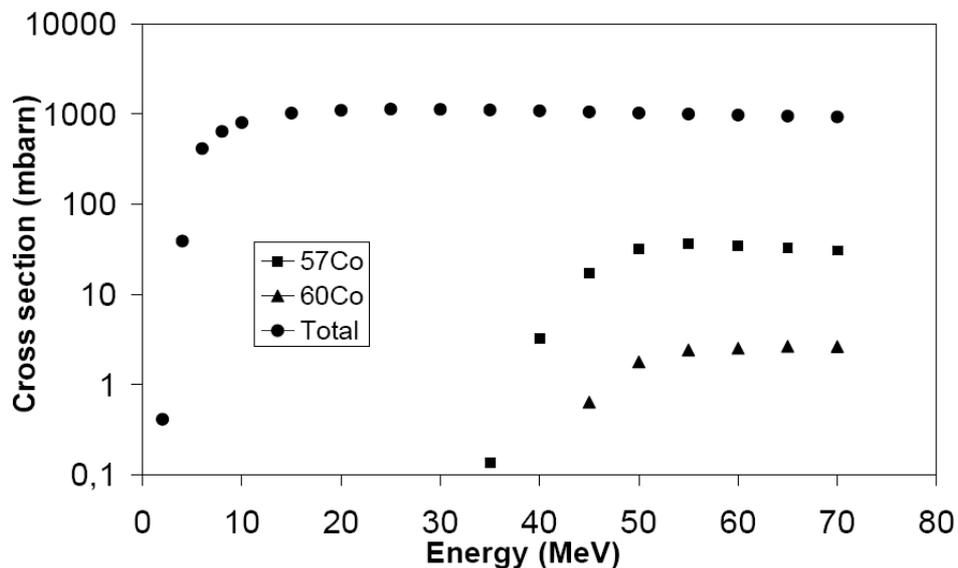

**Figure 8 Radionuclide production cross section from natural copper as a function of proton energy. Triangles are associated to $^{60}$Co production, square to $^{57}$Co and circles to the total reaction cross section.**

# Conclusion

An experimental setup is being installed at the ARRONAX facility to perform high energy PIXE. From our first measurement, we can expect ppm detection limit for medium mass isotopes as measured from a first experiment using a thick copper target. The detection limit is directly connected to the background. In order to reduce it, a new shielding of the detector is underway using three layers made of lead, copper and aluminium. Some modeling has started in order to get a better knowledge of the different background components. These calculations take into account electron bremsstrahlung and Compton diffusion. The next step will be to take into account the effect of secondary fluorescence in composed medium at high energy. Finally, we are also interested in measuring the K X-Ray production cross section for heavy elements these data will be necessary to take advantage of the high energy proton which are well suited to measure concentration profile of an element as a function of the depth in a thick target.

Gamma rays coming from nuclear reaction can be registered on-line and off-line. For thick sample studies, this additional information can substitute to low energy K X-rays which are greatly attenuate in matter. Thus gamma ray emission can help us by giving information of low Z isotopes deeper in the samples and can improve the detection limit. Sample activation, which is inherent to the use of high energy particles, should be manageable by either reducing the beam intensity, limiting the irradiation time or by selecting carefully particle incident energy in order to limit the production of long-lived isotopes.

## References

[1] K. Ishii and S. Morita, *Nucl. Instr. and Meth. B* **1988** *34*, 209.
[2] J. J. G. Durocher, N. M. Halden, F. C. Hawthorne, and J. S. C. McKee, *Nucl. Instr. And Meth. B* **1988** *30*, 470.
[3] J. L. Ruvalcaba and J. Miranda, *Nucl. Instr. and Meth. B* **1996** *109-110*, 121.
[4] H. Paul, *Nucl. Instr. and Meth. B* **1984** *3*, 5.
[5] N. M. Halden, *Nucl. Instr. and Meth. B* **1993** *77*, 399.
[6] A. Denker, W. Bohne, J. Opitz-Coutureau, J. Rauschenberg, J. Röhrich, and E. Strub, *Nucl. Instr. and Meth. B* **2005** *239*, 65.
[7] H. Homeyer, *Nucl. Instr. and Meth. B* **1998** *139*, 58.
[8] A. Denker and K. H. Maier, *Nucl. Instr. and Meth. B* **2000** *161-163*, 704.
[9] F. Haddad et al, *Eur. J. Med. Mol. Imaging* **2008** *35*, 1377-1387.
[10] K. Ishii, A. Yamadera, M. Sebata, and S. Morita, *Phys. Rev. A* 1981 *24*, 1720–1725.
[11] K. Ishii, *Radiat. Phys. Chem.* **2006** *75*, 1135.
[12] A.J. Koning, S. Hilaire and M.C. Duijvestijn, "TALYS-1.0", *Proceedings of the International Conference on Nuclear Data for Science and Technology - ND2007*, **2007** Nice France.
[13] C. Leroy, P. Rancoita, *Principles of Radiation Interaction in Matter and Detection*, World Scientific, Singapore, **2009**.